\documentclass[proof]{pasj01}

\begin{document}
\SetRunningHead{T. Sato et al.}{\textit{Suzaku} X-ray Observation of Kes 79}
\Received{}
\Accepted{}

\title{Suzaku spectra of a Type II Supernova Remnant Kes 79}

\author{Tamotsu \textsc{SATO},\altaffilmark{1,2}
Katsuji \textsc{KOYAMA},\altaffilmark{3,4}
Shiu-Hang \textsc{LEE},\altaffilmark{1}
and
Tadayuki \textsc{TAKAHASHI},\altaffilmark{1,2}}
\altaffiltext{1}{Department of High Energy Astrophysics, Institute of Space and Astronautical Science (ISAS), Japan Aerospace Exploration Agency (JAXA), 3-1-1Yoshinodai, Chuo, Sagamihara, Kanagawa, 252-5210, Japan}
\email{sato@astro.isas.jaxa.jp}
\altaffiltext{2}{Department of Physics, Graduate School of Science, University of Tokyo, Hongo 7-3-1, Bunkyo, Tokyo, 113-0033, Japan}
\altaffiltext{3}{Department of Earth and Space Science, Graduate School of Science, Osaka University, 1-1, Machikaneyama,Toyonaka, Osaka 560-0043, Japan}
\altaffiltext{4}{Department of Physics, Graduate School of Science, Kyoto University, Kita-Shirakawa, Sakyo-ku, Kyoto 606-8502, Japan}


\maketitle

\begin{abstract}
This paper reports results of a \textit{Suzaku} observation of the supernova remnant (SNR) Kes 79 (G33.6+0.1). The X-ray spectrum is best fitted by a two-temperature model: a non-equilibrium ionization (NEI) plasma and a collisional ionization equilibrium (CIE) plasma. The NEI plasma is spatially confined within the inner radio shell with $kT \sim$0.8~keV, while the CIE plasma is found in more spatially extended regions associated with the outer radio shell with $kT \sim$0.2~keV and solar abundance. 
Therefore, the NEI plasma is attributable to the SN ejecta and the CIE plasma is forward shocked interstellar medium. In the NEI plasma, we discovered K-shell line of Al, Ar and Ca for the first time.  The abundance pattern and estimated mass of the ejecta are consistent with the core-collapse supernova explosion of a $\sim$30--40~\Mo~progenitor star.  An Fe line with center energy of $\sim$6.4~keV is also found in the southeast (SE) portion of the SNR, a close peripheral region around dense molecular clouds.  One possibility is that the line is associated with the ejecta.  However, the centroid energy of $\sim$6.4~keV and the spatial distribution of enhancement near the SE peripheral do not favor this scenario. 
Since the $\sim$6.4~keV emitting region coincides to the molecular clouds, we propose another possibility that the Fe line is due to  K-shell ionization of neutral Fe by the interaction of locally accelerated protons (LECRp) with the surrounding molecular  cloud.  Both these possibilities, heated ejecta or LECRp origin, are discussed based on the observational facts.

This paper reports results of a Suzaku observation of the supernova remnant (SNR) Kes 79 (G33.6+0.1). The X-ray spectrum is best fitted by a two-temperature model: a non-equilibrium ionization (NEI) plasma and a collisional ionization equilibrium (CIE) plasma. The NEI plasma is spatially confined within the inner radio shell with kT~0.8 keV, while the CIE plasma is found in more spatially extended regions associated with the outer radio shell with kT~0.2 keV and solar abundance. 
Therefore, the NEI plasma is attributable to the SN ejecta and the CIE plasma is forward shocked interstellar medium. In the NEI plasma, we discovered K-shell line of Al, Ar and Ca for the first time.  The abundance pattern and estimated mass of the ejecta are consistent with the core-collapse supernova explosion of a ~30--40 solar mass progenitor star.  An Fe line with center energy of ~6.4 keV is also found in the southeast (SE) portion of the SNR, a close peripheral region around dense molecular clouds.  One possibility is that the line is associated with the ejecta.  However, the centroid energy of ~6.4 keV and the spatial distribution of enhancement near the SE peripheral do not favor this scenario. 
Since the ~6.4 keV emitting region coincides to the molecular clouds, we propose another possibility that the Fe line is due to  K-shell ionization of neutral Fe by the interaction of locally accelerated protons (LECRp) with the surrounding molecular  cloud.  Both these possibilities, heated ejecta or LECRp origin, are discussed based on the observational facts.

\end{abstract}

\section{Introduction}
\label{sec:introduction}

Kesteven 79 (or G33.6+0.1, hereafter Kes 79) is a Galactic supernova remnant (SNR) first discovered by Caswell et al. (1975) from analysis of radio sky surveys. Kes 79 is classified as a possible mixed-morphology SNR (Rho and Petre 1998). In the radio continuum, Very Large Array (VLA) observations have revealed a peculiar double-shell morphology of the SNR with two incomplete concentric rings of apparently different radii (Velusamy et al. 1991) whose origin is still under debate.  
The detections of a 1667~MHz OH absorption feature along the line-of-sight with $v_\mathrm{LSR} = +95$ to $+115$~km s$^{-1}$ (Green 1989) and  spatially coincident with a nearby molecular cloud (MC): the CO and HCO$^+$ emission at velocities in the same range (e.g., Scoville et al. 1987, Green and Dewdney 1992), suggest that the SNR shockwave is interacting with the dense molecular clouds.
The distance to the SNR has been estimated to be around 6.5 to 7.5~kpc through H\emissiontype{I} absorption (Case and Bhattacharya 1998) and 7.5~kpc by H\emissiontype{I} velocity measurements (Giacani et al. 2009). The apparent angular size thus implies a diameter of about 20~pc. 

Kes 79 is also known to be a GeV gamma-ray emitter from recent observations by the Fermi Large Area Telescope (LAT) (Auchettl et al. 2014). Considering the observed spectral shape and brightness of the GeV emission, the SNR is interacting with MCs; the gamma-rays are most naturally interpreted as having a `hadronic' origin; locally accelerated cosmic-ray (CR) protons collide with the surrounding dense clouds and produce neutral pions which subsequently decay into gamma-ray pairs. 

Tsunemi and Enoguchi (2002) reported that the X-ray spectrum of Kes 79 has a thermal origin and can be described by a non-equilibrium ionization (NEI) plasma with nearly  uniform temperature and abundances (sub-solar). Seward et al. (2003) discovered a central compact object (CCO) using Chandra, and suggested that the SNR originated from a core-collapse SN (CC-SNR). This SNR shows filamentary structures and faint emission near the southwest (SW) of the outer radio shell (Sun et al. 2004). 

X-ray spectra with better statistics were  obtained by XMM-Newton observations (Giacani et al. 2009). The global spectrum can be fitted with a single NEI plasma, where the abundances are near solar, except for Ar which shows an unusually large abundance of 5.0$^{+0.8}_{-1.6}$ solar.
However, all the previous observations lack sufficient photon statistics to separate the ejecta from the circumstellar components of the X-ray emission, which obviously makes it difficult to extract the mass and nature of the progenitor star.

Recently, Auchettl et al. (2014) compiled the accumulated data from XMM-Newton observations of Kes 79 and performed a spatially resolved spectral analysis.
They found that X-ray spectra from about half of the region require two plasma components:  a 0.24~keV PSHOCK model with solar abundances, and a $\sim$0.8~keV VPSHOCK with super solar abundances. They inferred that the former is emitted by the shocked ambient material (ISM), while the latter is due to shocked ejecta.
Although the spectral parameters of the VPSHOCK plasma component (i.e., temperatures, abundances and ionization parameters) are found to be roughly homogeneous in space, they claimed there is a hint of higher chemical abundances near the SNR center.

These previous spectral studies were performed for energies below 3--4~keV mainly due to the high  background in higher energy bands. In this paper, we report a new X-ray study of Kes 79 using \textit{Suzaku} data in the 0.7--7.5~keV band, taking advantages of the low and stable in-orbit background and the high sensitivity in the hard X-ray band of the XIS instrument (Mistuda et al. 2007, Koyama \etal\ 2007). We construct a model for the Galactic X-ray background emission to maximize the source photon statistics.  Fitting of the resulting source spectrum clearly requires a multi-temperature NEI plasma and  additional CIE plasma component. We found non-solar abundances in the NEI component with newly discovered K-shell lines of Al, Ar and Ca. The CIE plasma is close to solar abundance and is spatially extended toward the outer radio ring.  A clear Fe K-line near 6.4~keV from this CC-SNR is also detected for the first time.  We present our interpretations for the origin and structure of Kes79, particularly on the Fe-line emission. All errors in this report are 90\% confidence levels, unless otherwise specified.

\section{Observations and Data Reduction}
\label{sec:observation}

The data are obtained from the \textit{Suzaku} Key Project observations for the systematic X-ray study of Mixed-Morphology SNRs (PI=Koyama, K.).
We also use \textit{Suzaku} archival data near Kes 78, a SNR adjacent to Kes 79, to  estimate the local X-ray background. The observation log is summarized in Table~1.

\begin{table*}[htbp]
\caption{The Log of observations}
\label{tbl:obs}
\begin{center}
\begin{tabular}{cccc}
\hline
Sequence No. & observation galactic coordinate (l, b) & observation date & net exposure (ksec) \\ \hline
506059010 & (33.704, 0.0185) & 2011 April 23-25 & 50.0 \\
507036010 & (32.688, $-0.0762$) & 2012 April 21-22 & 52.2 \\ \hline
\end{tabular}
\end{center}
\end{table*}

The observations were performed by the X-ray Imaging Spectrometer (XIS: Koyama \etal\ 2007) onboard \textit{Suzaku}. The XIS instrument consists of four X-ray charge coupled device (CCD) cameras on the focal planes of the X-Ray Telescope (XRT: Serlemitsos \etal\ 2007). The XIS~0, 2 and 3 cameras have front-illuminated (FI) CCDs, whereas the XIS 1 is back-illuminated (BI). The XIS~2 camera has been out of operation since November 2006.

Data reduction is performed using the HEADAS software (version 6.16) and calibration database CALDB released on July 2014.
The spaced-row charge injection technique (Uchiyama et al. 2009) is applied to
compensate the degradation of energy resolution due to the charge transfer efficiency.
We exclude flickering and hot pixels, events collected during the South Atlantic Anomaly, as well as data with day and night-time earth elevation angles of less than 20$^\circ$ and 5$^\circ$ respectively. 
Data in the two editing modes (3$\times$3 and 5$\times$5 pixel events) are combined. After screening, the total exposure time is about 100~ks.
The redistribution matrix file (RMF) and the ancillary response file (ARF) for spectral analyses are generated by the xisrmfgen and xissimarfgen tools respectively (Ishisaki et al. 2007).

\section{Analysis and Results}
\label{sec:analysis}
\subsection{X-ray Image}

Figure~1 shows the XIS images in the energy bands of 0.7--1.5~keV and 2.0--4.0~keV. Here the X-ray events of XIS~0, 1 and 3 are combined, with corrections for the vignetting effect and subtraction of the non-X-ray background (NXB). The NXB is estimated by the xisnxbgen tool (Tawa \etal\ 2008).

\begin{figure}[htbp]
\begin{center}
\includegraphics[width=7cm]{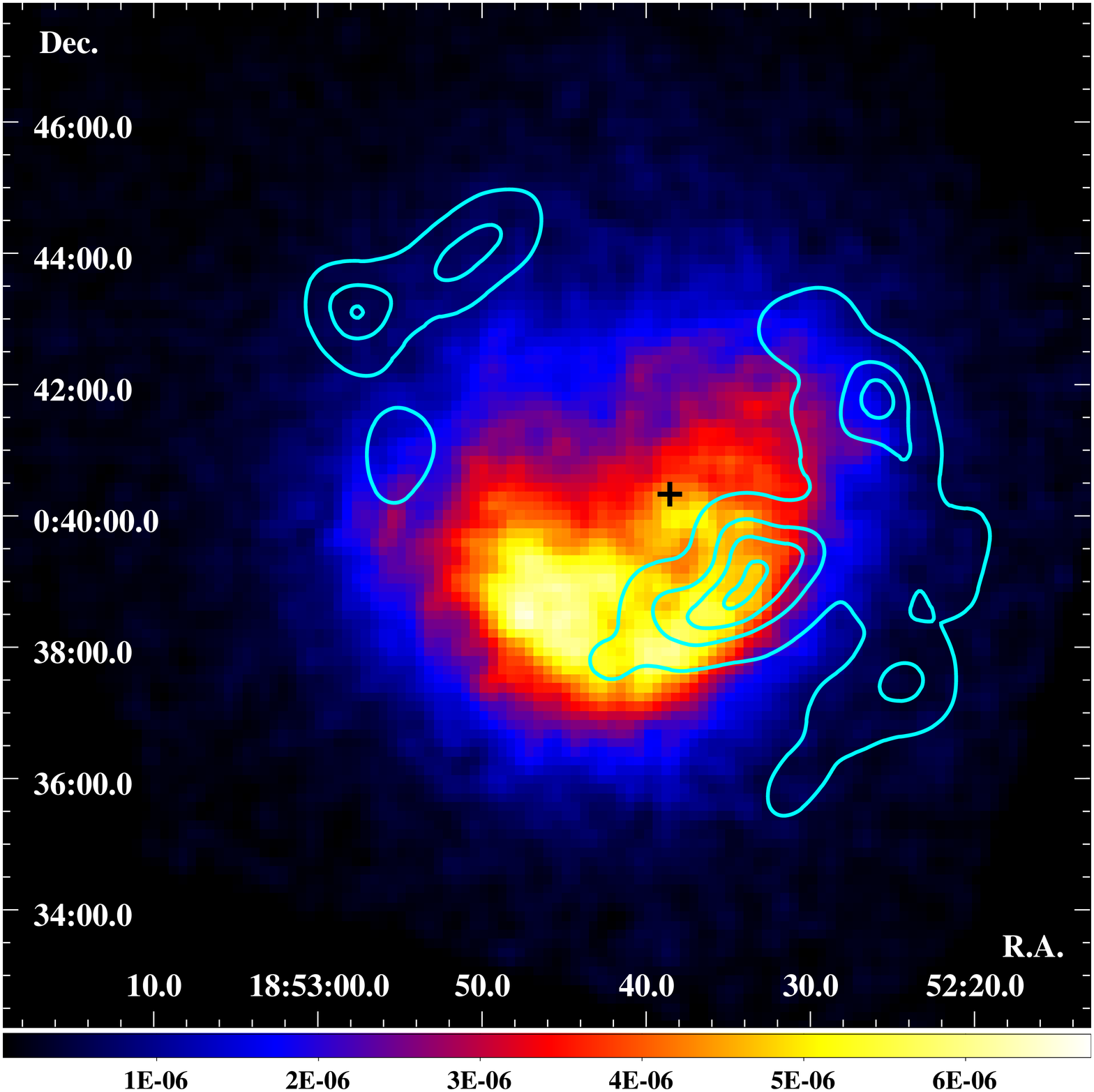}
\includegraphics[width=7cm]{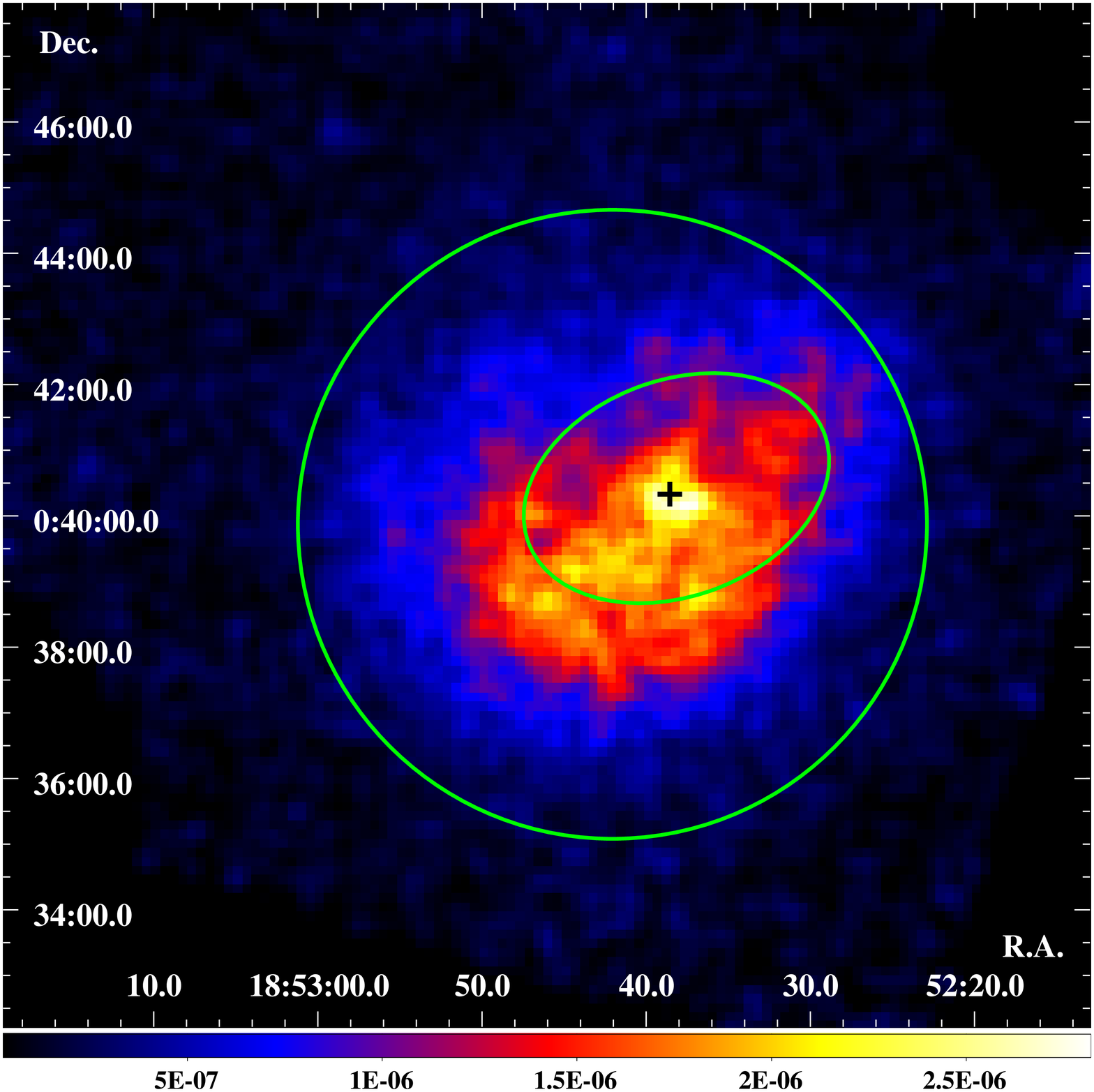}
\caption{
NXB-subtracted XIS images of Kes 79 after correcting for the vignetting effect in units of photons~s$^{-1}$~keV$^{-1}$~cm$^{-2}$. The  energy bands are 0.7--1.5~keV (left) and 2.0--4.0~keV (right). 
The location of the CCO is indicated by the black cross (Seward et al. 2003), and the cyan contours delineate the flux map of the 1.4 GHz emission (Condon et al. 1998). The green circles define the borders of the inner-ring and central regions of Kes 79.
}
\label{fig:image}
\end{center}
\end{figure}

The soft (0.7--1.5~keV) and hard (2--4~keV) band images clearly show contrasting morphologies; emission in the hard band (right) is more spatially concentrated around the CCO, while the softer emission (left) is more extended toward the SW of the outer radio shell (e.g. Giacani et al. 2009).
This morphological contrast strongly suggests that the X-ray emission of Kes 79 has at least two distinct plasma components. 

\subsection{Background Estimation}

The major contributor to the background of Kes 79 is the Galactic ridge X-ray emission (GRXE: Uchiyama \etal\ 2011, 2013), because the source location is on the Galactic plane at ($l, b)= (\timeform{33.6D}, \timeform{0.05D})$.  In order to constrain the GRXE with sufficient statistics and accuracy, we construct a background model using two blank skies: `BG1' near Kes 78 and `BG2' near Kes 79.
Figure~2 shows the X-ray images of the two backgrounds in the 0.8 to 3.0~keV band, where the two dashed lines define the BG1 and BG2 regions respectively.

\begin{figure}[htbp]
\begin{center}
\includegraphics[width=6cm]{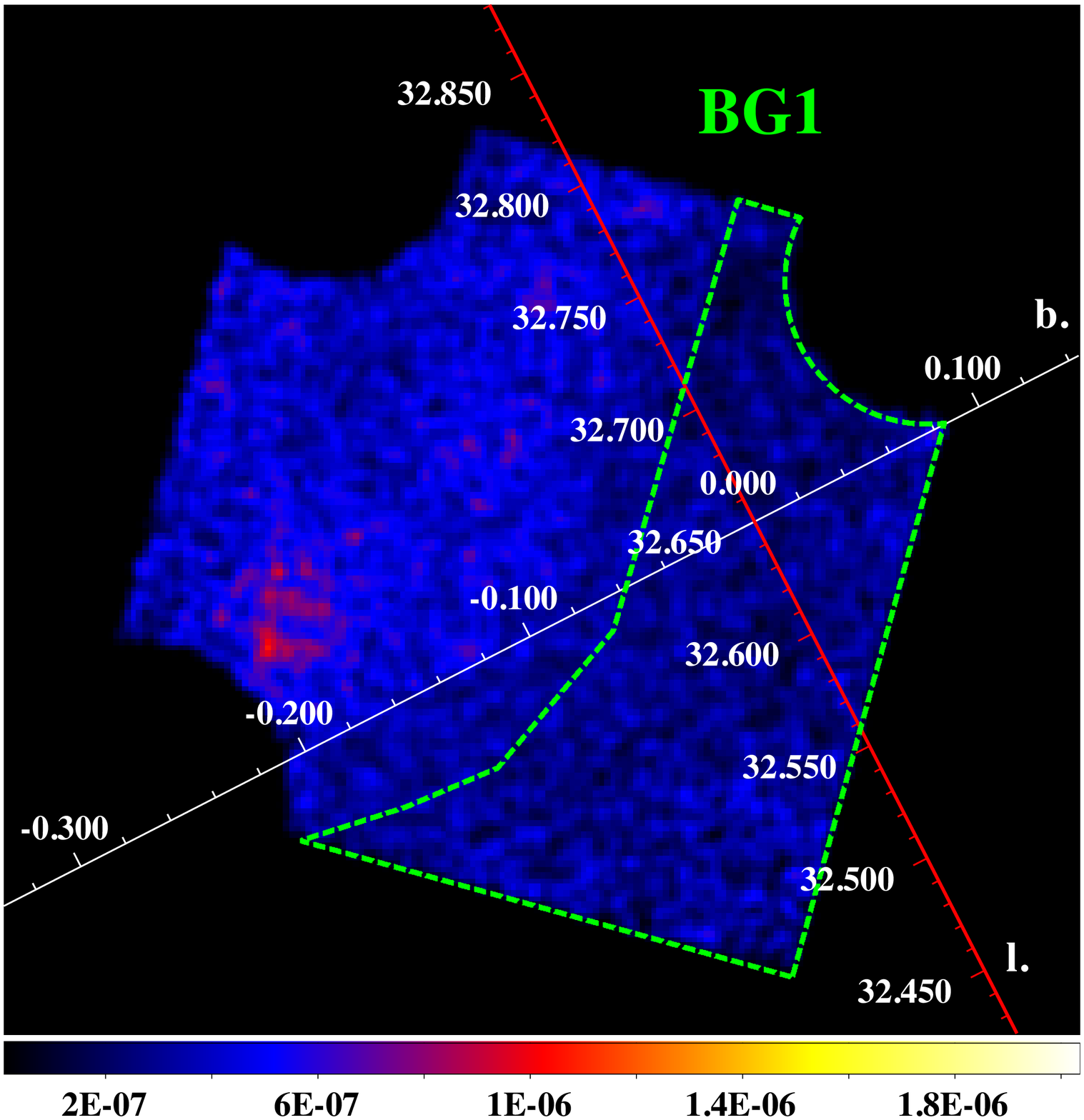}
\includegraphics[width=6cm]{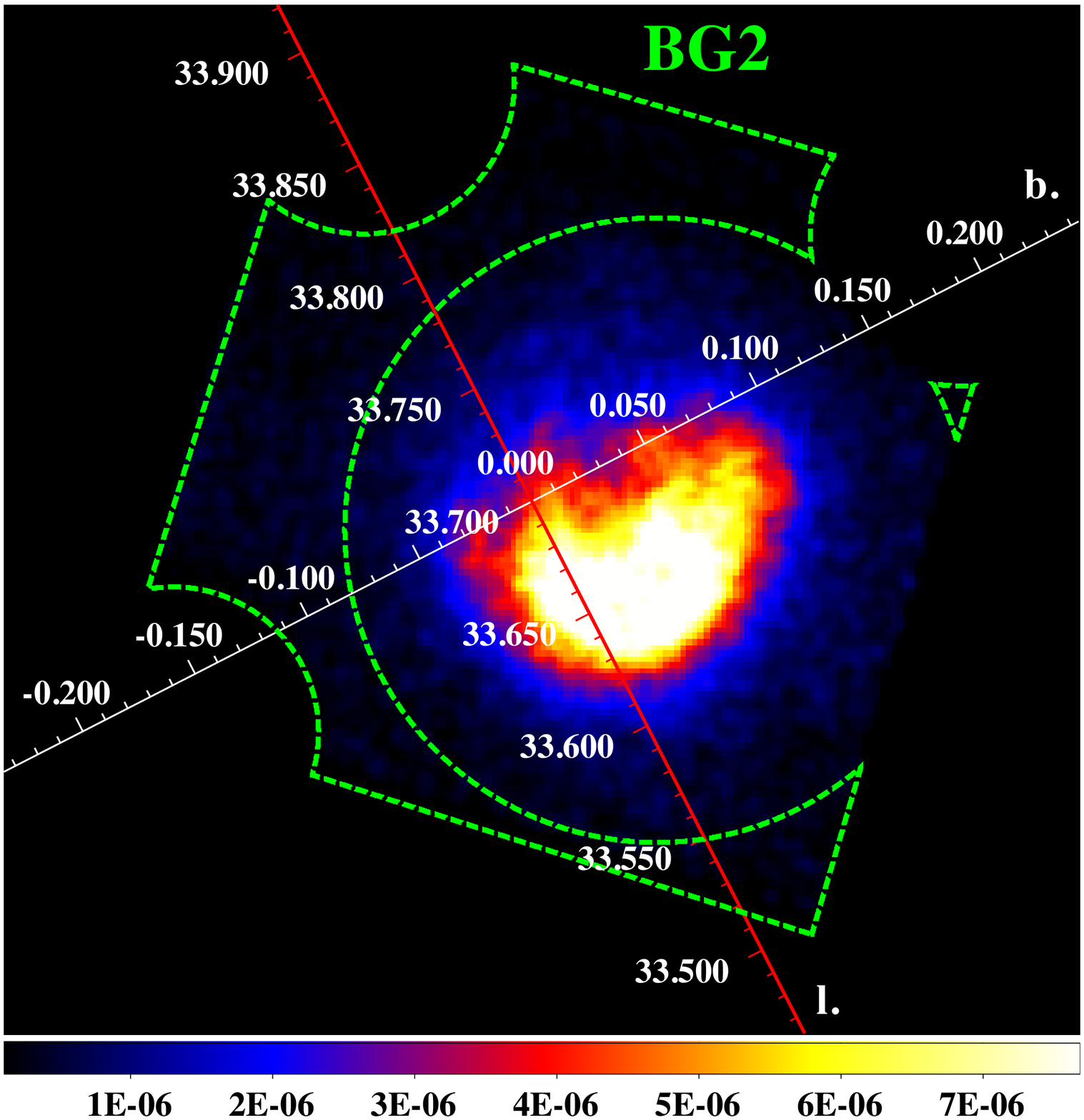}
\caption{
NXB-subtracted image (0.8--3.0~keV) of blank skies BG1 (left) and BG2 (right) after correcting for vignetting and exposure time in units of photons~s$^{-1}$~keV$^{-1}$~cm$^{-2}$. The dashed lines indicate the background regions. The red lines denote the galactic latitude $b=\timeform{0.0D}$.
}
\label{fig:image}
\end{center}
\end{figure}

For simplicity, we use the notation that the ``BG1-spectrum" is the NXB-subtracted spectrum from the region BG1, and similarly for BG2, Kes 79 (source), and so on (see next subsection for details). 
The BG1-spectrum and BG2-spectrum are mainly comprised of the Cosmic X-ray Background (CXB) and the Galactic ridge X-ray emission (GRXE).
The CXB has a power-law spectrum with a photon index of 1.41 and a flux of 8.2 $\times$ 10$^{-7}$~photons~cm$^{-2}$~s$^{-1}$~arcmin$^{-2}$~keV$^{-1}$ at 1~keV (Kushino \etal\ 2002). 
The GRXE consists of a low-temperature (LP) and a high-temperature (HP) CIE plasma, a cold matter (CM) emission with neutral iron line (6.4~keV) of equivalent width (EW) $\sim 450$~eV, and the foreground (FG) CIE plasma with a temperature of 0.59~keV (Uchiyama et al. 2013).
We ignore another FG plasma with a temperature of 0.09~keV, since it has no contribution to the relevant energy range of $>$0.7~keV.

Thus the BG1 and BG2-spectra can be modeled with the following algorithm:
\begin{equation}
A_{\mathrm{FG}} \times {\mathrm{FG}} + A_{\mathrm{GRXE}} \times ({\mathrm{LP + HP + CM}}) + 2A_{\mathrm{GRXE}} \times {\mathrm{CXB}}, 
\end{equation}
where $A_{\mathrm{FG}}$ and $A_{\mathrm{GRXE}}$ denote Galactic absorption with the column densities of $N_{\mathrm{H;FG}}$ for the foreground plasma and $N_{\mathrm{H;GRXE}}$ for the GRXE components, respectively. The absorption for the CXB is twice that for the GRXE because of the interstellar absorption by the front and back sides of the GRXE. 

To estimate the precise local GRXE, we use both BG1 and BG2 spectra.
Since the BG1 and BG2 regions are close to each other, their GRXE plasmas have nearly the same parameters.
However, the  BG2-spectrum is contaminated by the pile-over flux from Kes 79. In fact, the BG2-spectrum does show strong emission lines such as Mg\emissiontype{XI}~He$\alpha$ and Si\emissiontype{XII}~He$\alpha$ (Figure 3 right), which are attributable to contamination from Kes 79.
The contamination of BG1 by Kes 78, on the other hand, can be ignored because the Kes 78 flux is far lower than that of Kes 79.
Since BG2 is in the same XIS field as Kes 79, it provides us a more accurate background flux and absorption than those of BG1. To obtain the spectrum of the Kes 79 contamination, we determined phenomenological Kes79-spectra  by  subtracting the BG2-spectrum from the Kes79-spectrum (here Pheno-Kes79-spectrum).
We fit the Pheno-Kes79-spectrum with a 2-temperature bremsstrahlung model with Gaussian lines at K-shell energies of  Ne, Mg, Si, S, Ar and Fe, then obtained  reasonable fit with a $\chi^2$ value of $674.1/437\sim1.54$. The best-fit temperatures are $0.45\pm0.05$ and $1.28\pm0.11$~keV.

The BG1 and BG2-spectra are simultaneously fitted with the model given in Equation~(1), where the BG2-spectrum is added by $\sim$1.1\% of the flux from the best-fit Pheno-Kes79-spectrum. This contamination flux is estimated by using ray tracing (Ishisaki et al. 2007). 
The temperatures ($kT_{\mathrm HP}$) and abundances of Ne and Mg are treated as free parameters. The cross sections of interstellar photoelectric absorption are taken from Morrison and McCammon (1983).
The GRXE fluxes and absorption columns $N_{\mathrm{H}}$ for BG1 and BG2 are slightly different because of their different Galactic coordinates.  The scale length and height  ratios of BG2 and BG1 are calculated as $\sim$0.99 and $\sim$0.98, respectively.  Then the flux ratio of the GRXE (BG1/BG2) is estimated to be  0.97 (Uchiyama et al. 2011).
The simultaneous fit gives a reduced $\chi^2$ value of $673.80/567\sim1.19$.  
The best-fit spectrum and parameters are shown in Figure~3 and Table~2 respectively.
We use the GRXE parameters of BG2 around Kes 79 in the following spectral analysis. 

\begin{figure}[htbp]
\begin{center}
\includegraphics[width=7cm]{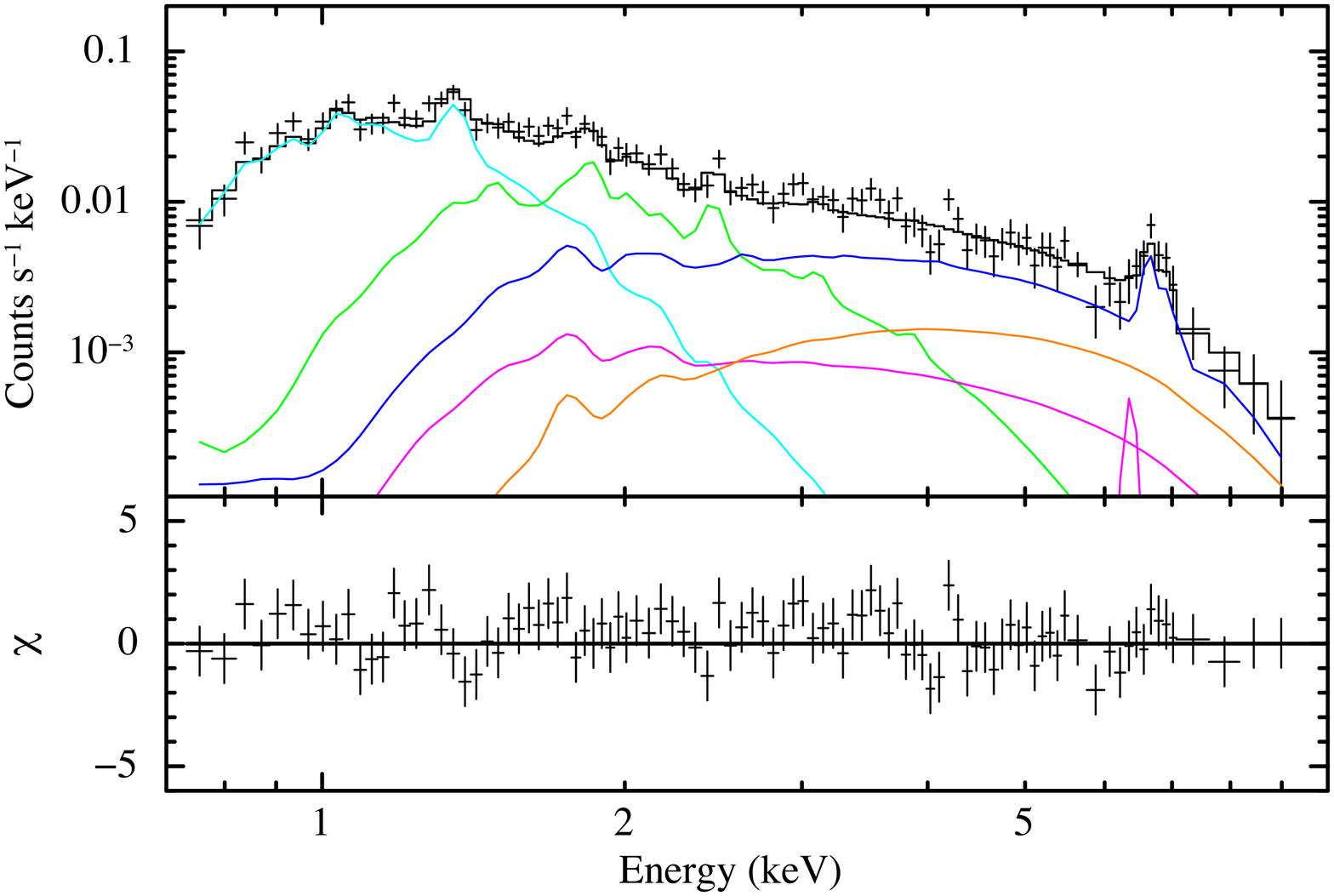}
\includegraphics[width=7cm]{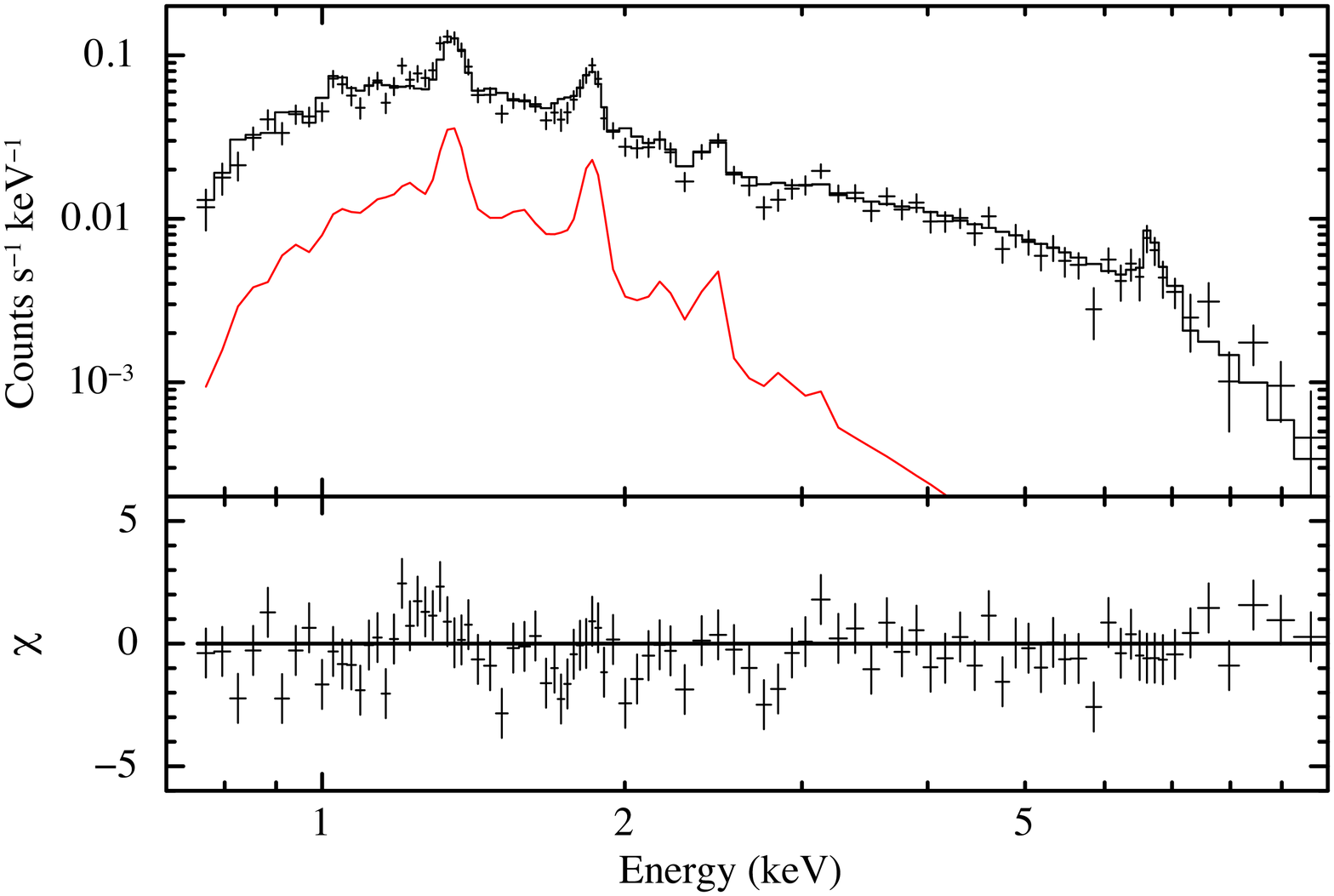}
\caption{Spectra extracted from the BG1 (left) and BG2 (right) blank skies. The best-fit models are shown by the solid black lines. For clarity, only the merged FI data are plotted. Left: the cyan, green and blue solid lines are the FG, LP and HP components respectively, whereas the magenta and orange lines show the Fe\emissiontype{I}~K$\alpha$ and CXB emission respectively.  Right: the red line shows the `leaked' flux from Kes 79 which we denote as the Pheno-Kes79-spectrum (see text).}
\label{fig:spec background}
\end{center}
\end{figure}

\begin{table}[htbp]
\caption{Fitting parameters of the background regions}
\begin{center}
\begin{tabular}{ccc}
\hline
\multicolumn{3}{c}{foreground} \\ \hline
$N_{\mathrm{H;FE}}$ (10$^{22}$~cm$^{-2}$) & \multicolumn{2}{c}{ $0.69\pm0.13$ } \\
$kT$ (keV) &  \multicolumn{2}{c}{ $0.39\pm0.04$} \\
Ne (solar) &  \multicolumn{2}{c}{ $0.13^{+0.07}_{-0.04}$ } \\
Mg (solar) &  \multicolumn{2}{c}{ $0.15^{+0.06}_{-0.03}$ } \\
others (solar) &  \multicolumn{2}{c}{ $0.05^{+0.05}_{-0.02}$ } \\ 
Flux ($10^{-6}$~s$^{-1}$~cm$^{-2}$~arcmin$^{-2}$)& \multicolumn{2}{c}{ $1.4^{+0.5}_{-0.6}$ (BG1), $1.5^{+0.5}_{-0.4}$(BG2) (0.7--1.2~keV)} \\ \hline
 GRXE & LP & HP \\ \hline
$N_{\mathrm{H;GRXE}}$~(10$^{22}$~cm$^{-2}$) & \multicolumn{2}{c}{ $2.5\pm0.4$ (BG1), $2.9^{+0.4}_{-0.5}$(BG2)} \\
$kT$ (keV) & $0.90^{+0.08}_{-0.10}$ & $8.4^{+1.8}_{-1.6}$ \\
Abundance (solar) & $0.44^{+0.24}_{-0.15}$ & $1.1^{+0.7}_{-0.4}$ \\
Flux (BG1) ($10^{-6}$~cm$^{-2}$~s$^{-1}$~arcmin$^{-2}$) & \multicolumn{2}{c}{1.2$^{+0.7}_{-0.4}$ (2.3--5~keV), $0.42\pm0.13$ (5--8~keV)} \\
Fe\emissiontype{I}~K$\alpha$ ($10^{-8}$~cm$^{-2}$~s$^{-1}$~arcmin$^{-2}$) & \multicolumn{2}{c}{$0.8^{+1.1}_{-0.8}$ } \\
Flux ratio of BG2/BG1 & \multicolumn{2}{c}{ 0.97 (fix) }  \\ \hline
$\chi^2$ / d.o.f & \multicolumn{2}{c}{ 673.80/567 } \\ \hline
\end{tabular}
\end{center}
\label{tbl:par background}
\end{table}

\subsection{The plasma model fit to the overall spectrum of Kes 79}

The overall spectrum of Kes 79 (Kes79-spectrum) is extracted from the outer green circle in Figure~1 (right panel), after subtracting the NXB.
We fit the Kes79-spectrum with a thermal plasma model adding the BG2-spectrum (see Table~2) and the spectral model of the CCO from Seward (2003). 
We use the plasma code VVNEI in the XSPEC package to represent a non-equilibrium ionization (NEI) plasma.  The electron temperature ($kT$) and ionization parameter ($n_{\mathrm e} t$) are free parameters, where $n_{\mathrm e}$ and $t$ are the electron density and elapsed time following shock-heating.  The abundances of Ne, Mg, Al, Si, S, Ar, Ca and Fe are free parameters, and those of the other elements are fixed to 1 in units of solar abundance.
This model, however, leaves a line-like residual at 1.2~keV, which corresponds to the higher Rydberg series of Fe-L. In fact, the atomic data for the Fe-L line complex around 1.2~keV are incomplete in the current plasma codes.  We therefore add a Gaussian line with a free center energy and flux. 

Since this model is still unacceptable with a reduced $\chi^2$ / d.o.f $\sim$1.38 with a clear excess at low energies, we add a low-temperature CIE plasma component (APEC; Smith et al. 2001). The best-fit abundances of the APEC plasma are nearly solar. We therefore fix all the abundances to solar.
The fit of the two-temperature NEIs plus CIE model is poor with a reduced $\chi^2$ / d.o.f  of $\sim$1.18.  Significant residuals are still seen at 6.4~keV and around 2--3~keV (Figure 4, middle panel).  

\begin{figure*}[htbp]
\begin{center}
\includegraphics[width=10cm]{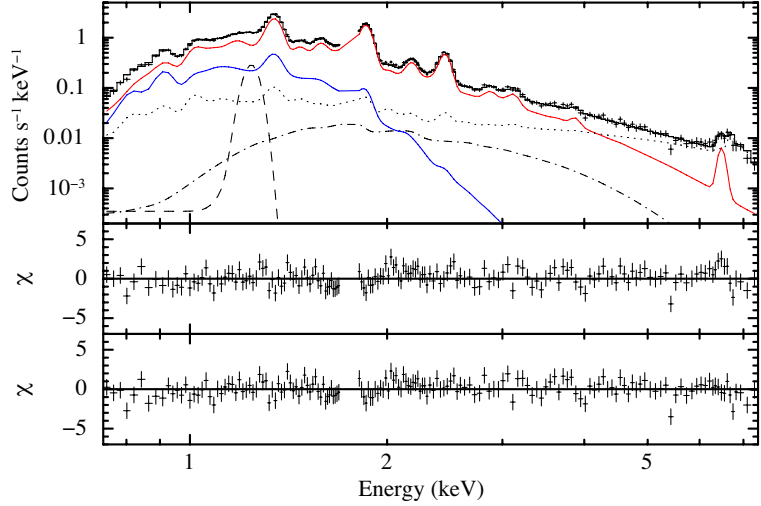} 
\caption{Top panel: The best-fit spectral model (solid line) for Kes 79. The multi-temperature NEI and CIE components are shown by the red and blue lines. The 1.2~keV Gaussian line, BG and CCO components are shown by the dotted, dashed and dash-dotted lines.
For clarity, only the merged FI spectrum is plotted. Middle panel: residuals from the NEI + CIE + 1.2~keV Gaussian line model. Bottom panel: residuals from the multi-temperature NEI + CIE + 1.2~keV line model.}
\label{fig:spec}
\end{center}
\end{figure*}

The residual at 6.4~keV is due to the K-shell line of nearly neutral Fe. 
Figure~5 is an expanded version of the Kes 79 and BG1+BG2 spectra around 6.4~keV.
The solid lines are the  best-fit model of bremsstrahlung emission plus Gaussian lines. The best-fit Gaussian line of the Kes79-spectrum at $\sim$ 6.4~keV has a flux of $4.1\times 10^{-6}$~cm$^{-2}$~s$^{-1}$, while that of the BG1+BG2 spectrum is $5.6\times 10^{-7}$~cm$^{-2}$~s$^{-1}$. 
The maximum uncertainty from background subtraction is less than $\sim10$\% or $5.6\times 10^{-8}$~cm$^{-2}$~s$^{-1}$. This value is only $\sim2$\% of the 6.4~keV line flux detected in the Kes79-spectrum. Hence, we conclude that the detection of the 6.4~keV line feature is robust. 

\begin{figure*}[htbp]
\begin{center}
\includegraphics[width=7cm]{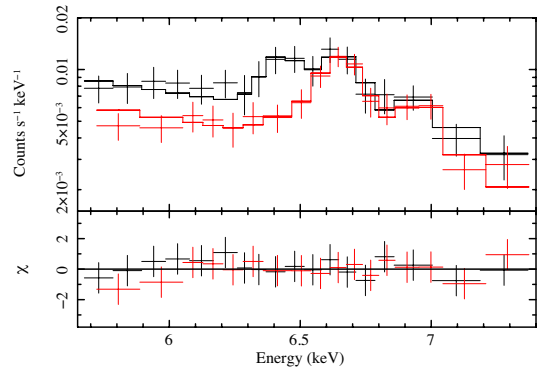}
\caption{The Kes79-spectrum (black), and the BG1+BG2 spectrum (red) used for background subtraction, zoomed in around 6.4~keV. The solid lines represent our best-fit models which include a bremsstrahlung component plus Gaussian lines.
}
\label{fig:spec}
\end{center}
\end{figure*}
 
The excess at round 2--3~keV is due to an underestimation of the K-shell line complexes from Si and S.  For further improvement, we fit the temperatures of the H--Mg, Si--Ca and Fe--Ni elemental groups in the NEI plasma as independent free parameters, while  the normalizations are linked to each other. The best-fit $\chi^2$/d.o.f. is improved to $\sim$1.09 (Figure 4, lower panel).
This improvement is significant with a F-test null probability better than $\sim$4~\%. 

Statistically, this model is still unacceptable in the $\chi^2$ test, possibly due to some systematic errors.  In fact, we can obtain no further improvement by adding more components and/or changing to other trial models. 
We therefore adopt this model as the  best approximation for the overall spectrum from Kes 79. The best-fit model and spectral parameters are given in Figure 4 and Table 3 (Model A), respectively. In this model (Model A), the 6.4~keV line is regarded as the ejecta plasma with low ionization of Fe. 

The observed centroid energy of the 6.4 keV line does not exclude an origin from neutral irons, thus it is probable that the line is not emerging from a hot plasma in an extremely low ionization state.  We therefore attempt to fit this line by a Gaussian line (Model B).  In Model B, the plasma component is essentially the same as Model A, but the NEI plasma now consists of two temperature components for the H--Al and Si--Fe elemental groups respectively. Model B yields an equally good fit as Model A, with a $\chi^2_{\nu}/{\rm d.o.f.}$ of $\sim$1.10. The best-fit parameters are given in Table 3 (Model B).

\begin{table}[htbp]
\caption{Best-fit parameters of multi-kT+ CIE + Gaussian line model}
\begin{center}
\begin{tabular}{ccc}
\hline\hline
& \multicolumn{2}{c}{Overall} 		\\
& model A & model B \\ \hline
$N_{\rm H}$(10$^{22}$cm$^{-2}$)  & $1.44^{+0.03}_{-0.04}$& $1.49\pm0.03$ \\ \hline
\multicolumn{3}{c}{NEI}\\
\hline
$kT_1$ (keV) 	&$0.81^{+0.06}_{-0.05}$&$0.84\pm0.05$  \\
$n_{\rm e}t_1$ (10$^{10}$cm$^{-3}$~s)  	&$6.8\pm0.8$	& $6.3^{+1.1}_{-0.9}$ \\
Ne 		&$0.92\pm0.13$&$1.02^{+0.13}_{-0.12}$	 \\
Mg	&$1.12^{+0.08}_{-0.10}$&$1.19^{+0.10}_{-0.05}$  \\
Al &$0.84^{+0.15}_{-0.18}$&$0.8\pm0.2$		\\ \hline
$kT_2$ (keV)  		&$0.96^{+0.11}_{-0.10}$&$0.95^{+0.16}_{-0.12}$ \\
$n_{\rm e}t_2$ (10$^{10}$cm$^{-3}$~s) 	&$6.0^{+1.7}_{-1.3}$ &$4.7^{+1.3}_{-1.1}$ 	 \\
Si	& 	$0.82^{+0.15}_{-0.12}$& 	$0.88^{+0.19}_{-0.16}$	\\
S	& 	$0.93^{+0.20}_{-0.15}$ & 	$1.09^{+0.26}_{-0.22}$	\\
Ar & $0.7^{+0.3}_{-0.2}$ 	& $1.0^{+0.4}_{-0.3}$ \\
Ca   & $1.7^{+1.6}_{-1.0}$ & $2.8^{+2.3}_{-1.6}$ 	\\ \hline
$kT_{\rm 3}$ (keV)& $2.5^{+0.3}_{-0.2}$ 	& $= kT_2$	\\
$n_{\rm e}t_3$ (10$^{10}$~cm$^{-3}$~s) & $1.5\pm0.2$& $= n_{\rm e}t_2$\\
Fe = Ni & $0.35\pm0.05$& $0.66^{+0.16}_{-0.13}$			\\
$EM^\dag$ ($10^{12}$cm$^{-5}$) & $3.8\pm0.6$ & 3.6$\pm$0.5 	\\ \hline
Fe K$\alpha$ energy  & - & 6.39--6.48~keV \\
flux ($10^{-6}$~cm$^{-2}$~s$^{-1}$)  & - & $3.0\pm1.1$\\ \hline
\multicolumn{3}{c}{APEC}\\
\hline
$kT_{\rm e}$ (keV) 	 &$0.22^{+0.02}_{-0.03}$	&$0.22^{+0.02}_{-0.03}$	\\
Abundances& 1 (fixed)  & 1 (fixed)\\
$EM^\dag$ ($10^{12}$cm$^{-5}$) & $20^{+9}_{-7}$ &  $21^{+11}_{-3}$ \\
\hline
$\chi^2_{\nu}/{\rm d.o.f.}$ & 493.15/451& 497.43/451  \\
\hline
\multicolumn{3}{l} {\small $^*$$n_{\rm e}$$n_{\rm p}$$V$/4$\pi$$D^2$,where $n_{\rm e}$, $n_{\rm p}$, $D$ and $V$ are the electron }\\[-1.5mm]
\multicolumn{3}{l} {\small and proton number densities (cm$^{-3}$),the distance (cm)}\\[-1.5mm]
\multicolumn{3}{l} {\small and the emission volume (cm$^3$), respectively.}
\end{tabular}
\end{center}
\end{table}

\subsection{Spatially Resolved Spectra}

Auchettl et al. (2014) reported that the spatial distribution of temperature and abundances in Kes 79 is roughly uniform, but show a hint of small enhancement in the central region.  We therefore divide the whole region into two parts: the central region around the CCO with a size of $5\times3$~arcmin$^2$ (i.e., the region bounded by the inner green ellipse in the right panel of Figure~1), and the inner radio shell region (the remaining region of Kes 79 in the right panel of Figure~1).  We perform spectral analyses for the spectra from each regions with the same procedures and the same model (Model A and Model B) as are given in Section 3.3. 
In the center region spectrum, the abundances of Ca and Fe cannot be constrained. We therefore re-fit this spectrum by linking the abundance of Ca to that of Ar. For the same reason, the temperature and ionization parameter of the Fe--Ni plasma are linked to those of the Si--Ca plasma. We obtained reasonable fits with both Model A and B. The best-fit spectra and parameters for Model B are summarized in Figure~6 and Table~4. Since Model A gives essentially the same parameters value of Model B, we list only the Model B results. 

\begin{figure*}[htbp]	
\begin{center}
\includegraphics[width=7cm]{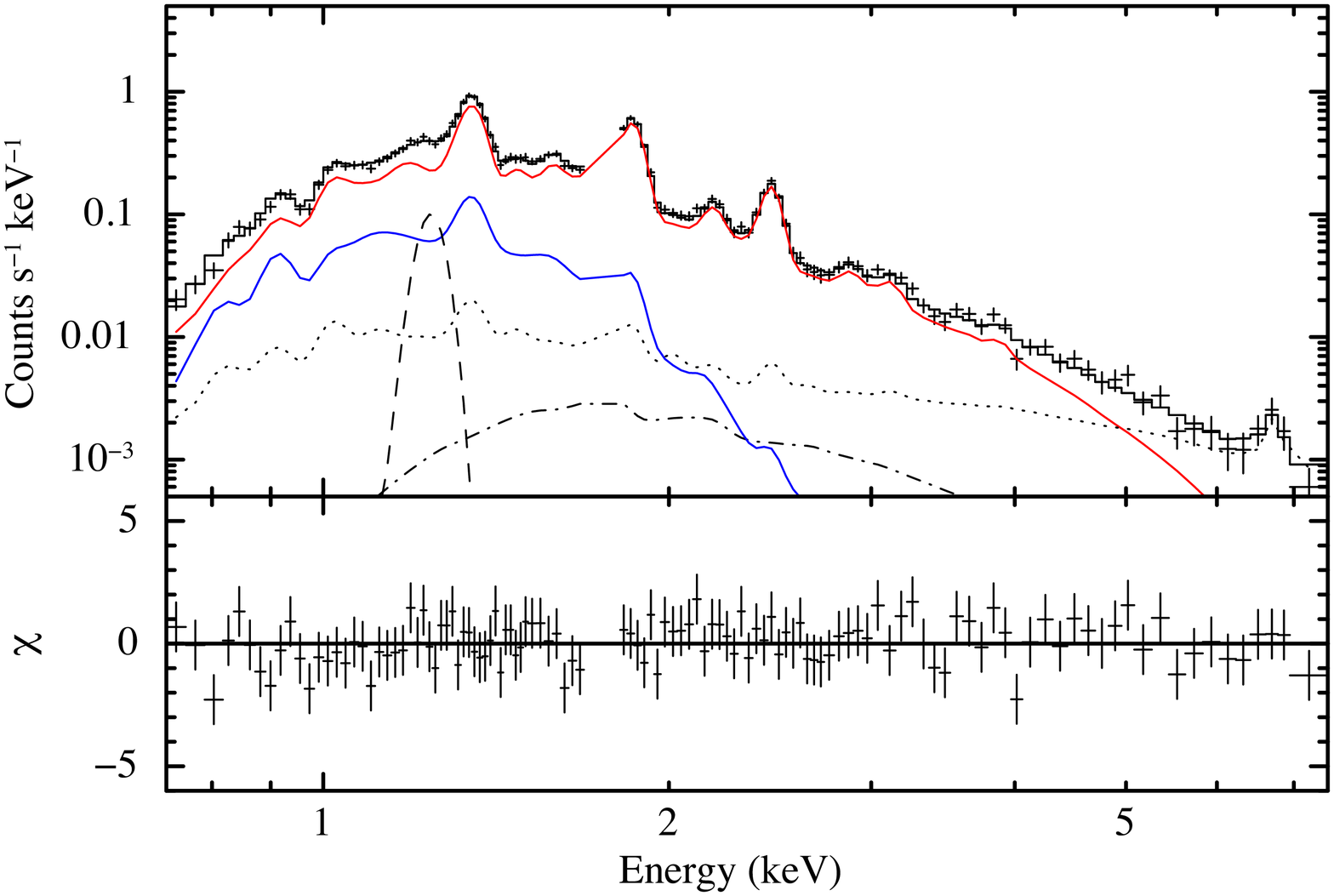} 
\includegraphics[width=7cm]{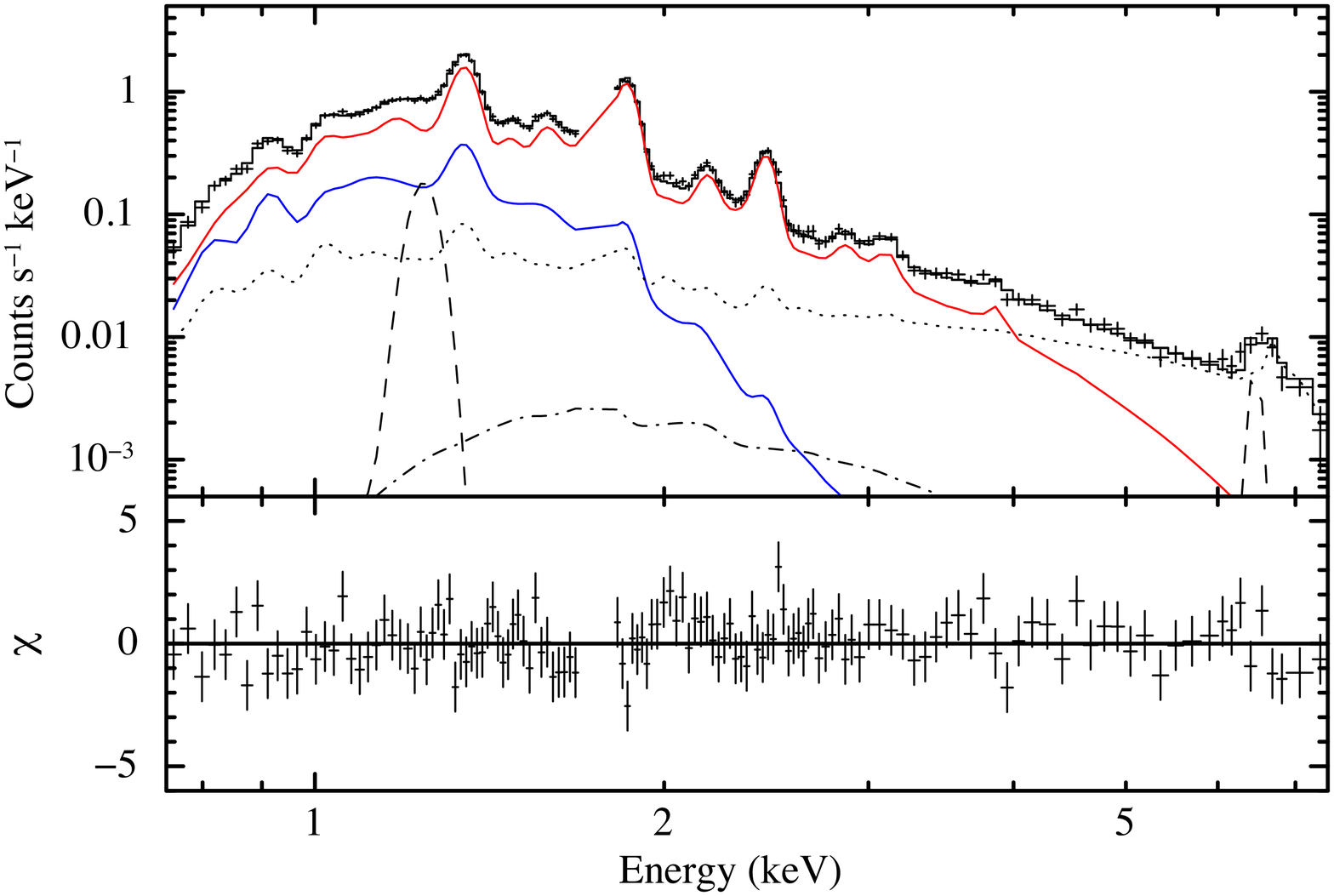} 
\caption{Spectra extracted from the center region (left) and the inner-ring region (right). The best fit models are shown as the solid lines (model B).
The multi-temperature NEI and CIE components are shown by the red and blue lines. The 1.2~keV and 6.4~keV Gaussian lines, BG and CCO components are shown by the dotted, dashed and dash-dotted lines.
For clarity, only the merged FI spectrum is plotted.}
\label{fig:spec}
\end{center}
\end{figure*}

\begin{table}[htbp]
\caption{Best-fit parameters of spatially resolved spectra.}
\begin{center}
\begin{tabular}{ccc}
\hline\hline
&  Center 		& Inner-Ring\\  \hline
$N_{\rm H}$(10$^{22}$cm$^{-2}$) & $1.54^{+0.06}_{-0.07}$	& $1.47\pm0.04$\\ 
\hline
\multicolumn{3}{c}{NEI}\\
\hline
$kT_1$ (keV) &$0.82^{+0.06}_{-0.05}$		&$0.83\pm0.06$\\
$n_{\rm e}t_1$ (10$^{10}$cm$^{-3}$~s)  &$6.9\pm1.4$ 		&$6.3\pm1.2$ 	 \\
Ne & $0.86^{+0.20}_{-0.08}$ 		& $1.01^{+0.17}_{-0.15}$ \\
Mg& $0.91^{+0.11}_{-0.10}$ 		&$1.25^{+0.13}_{-0.11}$ 	 \\
Al&$0.4\pm0.3$ 		&$1.0\pm0.3$\\ 
\hline
$kT_2$ (keV)&$1.0^{+0.3}_{-0.2}$		&$0.99^{+0.19}_{-0.13}$\\
$n_{\rm e}t_2$ (10$^{10}$cm$^{-3}$~s)&$5^{+4}_{-2}$ 		&$4.4^{+1.7}_{-1.4}$ 	 \\
Si&$0.6\pm0.2$ 	&$0.9^{+0.3}_{-0.2}$ \\
S&$0.8^{+0.4}_{-0.3}$	 	&$1.0\pm0.3$ \\
Ar& $0.5^{+0.4}_{-0.3}$	 	& $1.0^{+0.4}_{-0.3}$ \\
Ca & =Ar	 	& $3.0^{+1.9}_{1.7}$ \\ \hline
$kT_{\rm 3}$ (keV)&$= kT_2$ 	& $= kT_2$	\\
$n_{\rm e}t_3$ (10$^{10}$~cm$^{-3}$~s) & $= n_{\rm e}t_2$& $= n_{\rm e}t_2$\\
Fe = Ni & $0.4\pm0.2$ 	& $0.8\pm0.2$ \\
$EM^\dag$ ($10^{12}$cm$^{-5}$) &$1.5\pm0.3$ 	& $2.3\pm0.4$ \\ \hline
Fe K$\alpha$ energy  & 6.4~keV (fix) & 6.39--6.50~keV \\
flux ($10^{-6}$~cm$^{-2}$~s$^{-1}$)  & $<0.5$ & $2.8\pm1.0$\\
\hline
\multicolumn{3}{c}{APEC}\\
\hline
$kT_{\rm e}$ (keV) & $0.22\pm0.05$	& $0.22^{+0.02}_{-0.03}$\\
Abundances& 1 (fixed)& 1 (fixed) \\
$EM^\dag$ ($10^{12}$cm$^{-5}$) & $5^{+8}_{-3}$ 	& $14^{+9}_{-5}$ \\ \hline
$\chi^2_{\nu}/{\rm d.o.f.}$ & 278.30/349 	&415.60/362 \\
\hline
\multicolumn{3}{l} {\small $^*$$n_{\rm e}$$n_{\rm p}$$V$/4$\pi$$D^2$,where $n_{\rm e}$, $n_{\rm p}$, $D$ and $V$ are the electron }\\[-1.5mm]
\multicolumn{3}{l} {\small and proton number densities (cm$^{-3}$),the distance (cm)}\\[-1.5mm]
\multicolumn{3}{l} {\small and the emission volume (cm$^3$), respectively.}
\end{tabular}
\end{center}
\end{table}

In Table 4, we see no significant difference in temperatures and ionization timescales between the two regions for the H--Ca plasma.
We can see small enhancements of the metal abundances in the inner-ring region compared the center region. The normalization ratio between the CIE and NEI components (CIE/NEI) of the inner-ring region ($\sim6$) is about twice as large as that of the center region ($\sim3$).

Our analysis of spatially resolved spectra unveils a remarkable contrast in the distribution of the 6.4 keV iron line.
The 6.4~keV line is found in the inner-ring region, but not in the center region around the CCO (Table 4). The spatial distribution of the 6.4~keV line is examined  by constructing a narrow band image over the  6.2--6.5~keV energy band, shown in Figure~7. In the image the XIS 1 data are not included, because the BI CCD has much poorer signal-to-noise ratios at high energies around 6.4 keV. The CCO and the outline of the inner radio shell are shown by the red cross and cyan circle, respectively. The magenta contours show the $^{13}$CO $J = 1\rightarrow0$ emission integrated between +99.0 and +109.0~km~s$^{-1}$ (Giacani et al. 2009). This figure clearly show that the 6.4~keV line is concentrated at the dense MC region. 

\begin{figure}[htbp]
\begin{center}
\includegraphics[width=8cm]{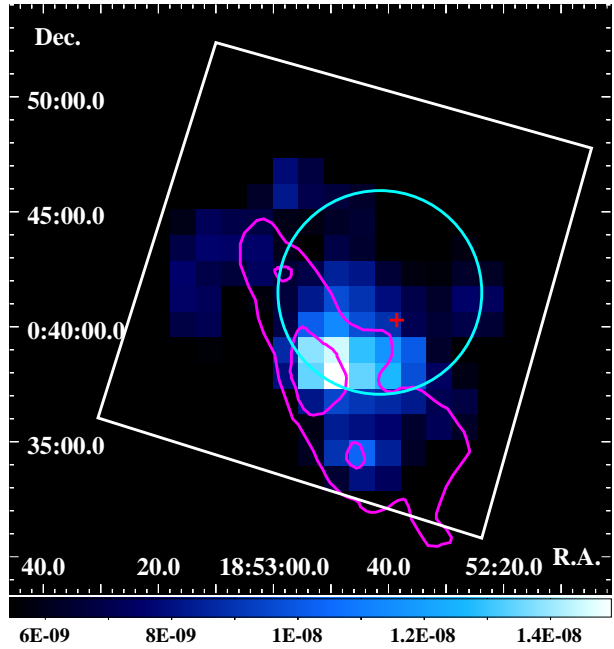}
\caption{The NXB-subtracted XIS 0+3 image of Kes 79 in the energy band of 6.2--6.5 keV after correction for vignetting  in units of photons~s$^{-1}$~keV$^{-1}$~cm$^{-2}$. The white square shows the FOV of the XIS. The CCO and the inner radio shell are indicated by the red cross and cyan circle (Seward et al. 2003, Sun et al. 2004). The magenta contours indicate the $^{13}$CO $J = 1\rightarrow0$ line emission integrated between +99.0 and +109.0~km~s$^{-1}$ (Giacani et al. 2009).}
\label{fig:spec}
\end{center}
\end{figure}

\subsection{X-ray Emission from the Outer Radio Ring}

Sun et al. (2004) found that Kes 79 has extended X-ray emission in the southwest (SW) of the outer radio shell.
We examined  more extended X-ray emission around the whole outer radio shell.
The surface brightness is very low and a hint of Ne He$\alpha$ line at $\sim$0.9 keV is found from the outer radio ring. We therefore extract and display a 0.86--0.96~keV band image in Figure~8 (left panel). Then we divided the outer ring into 4 azimuthal segments (NE: red, NW: magenta, SE: green and SW: orange) and made radial profiles as shown in Figure~8 (right top panel). To illustrate the results more clearly, 
the radial profiles from all segments are summed together for the Ne~\emissiontype{IX}~He$\alpha$ band (0.86--0.96~keV) and the Si~\emissiontype{XIII}~He$\alpha$ (1.76--1.92~keV) band respectively, and their ratios are plotted in Figure 8 (right bottom panel).

\begin{figure}[htbp]
\begin{center}
\includegraphics[width=8cm]{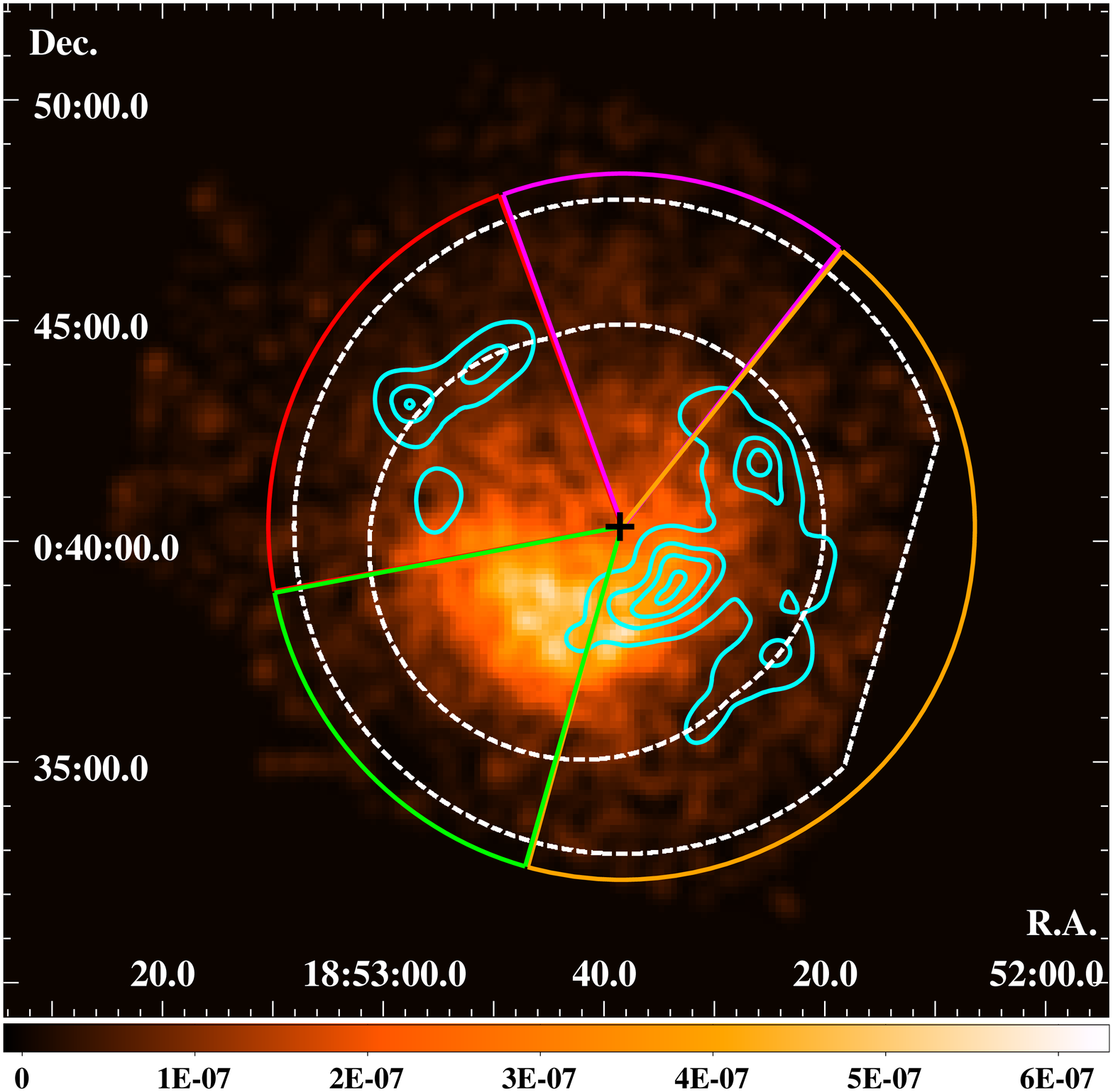}
\includegraphics[width=8.5cm]{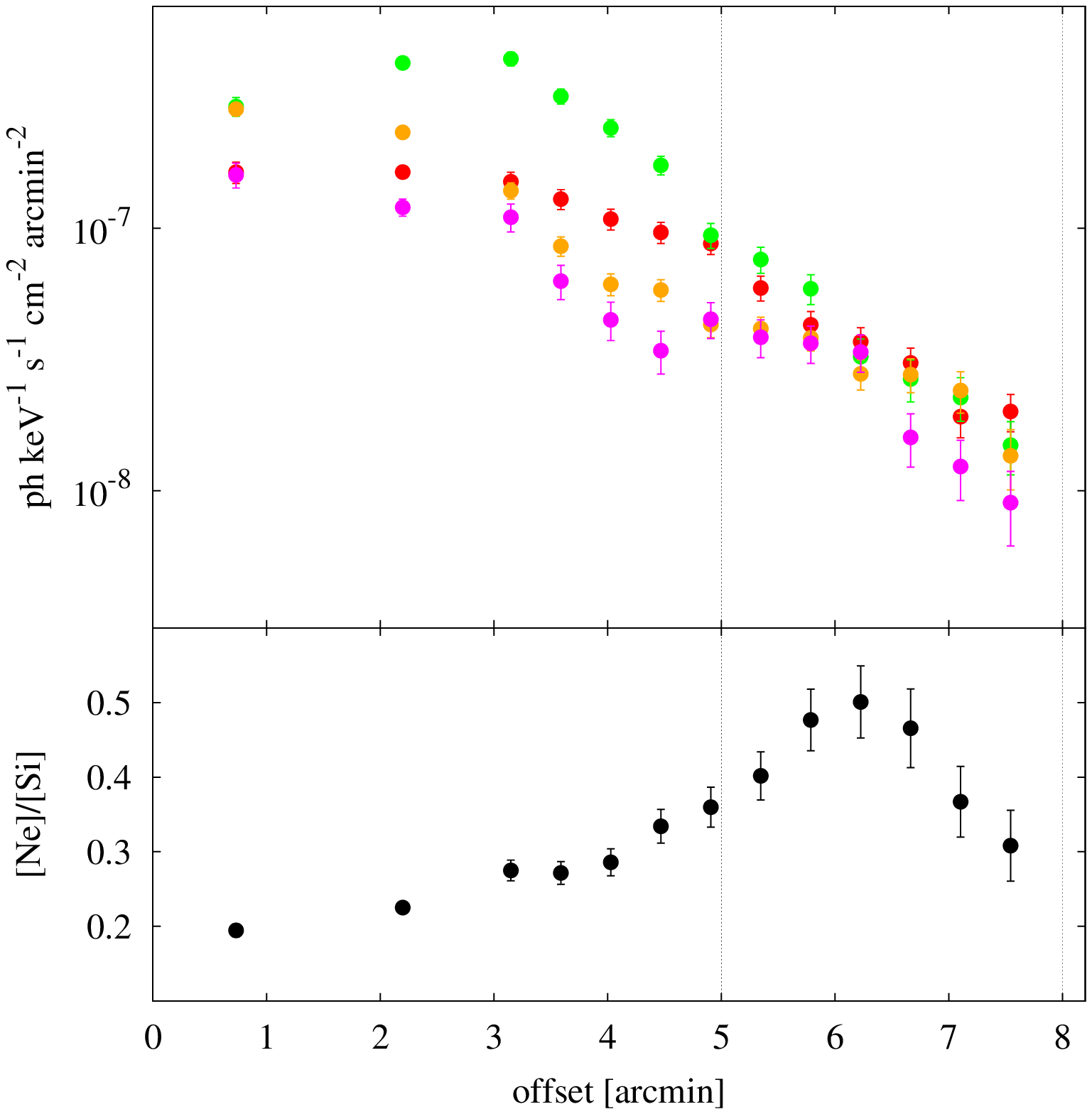}
\caption{Left : NXB-subtracted X-ray image (0.86--0.96~keV) of Kes 79 after correction for vignetting  in units of photons~s$^{-1}$~keV$^{-1}$~cm$^{-2}$. The location of the CCO is shown by the black cross (Seward et al. 2003). The image is smoothed by a Gaussian kernel with $\sigma = \timeform{1.3'}$. The colored sections correspond to our 4 azimuthal segments. The outer-ring region is bounded between the two white dashed lines.
Right (top) : Ne~\emissiontype{IX}~He$\alpha$ (0.86--0.96~keV) radial profiles from the 4 segments, with the angular distance (offset) from the CCO position. Right (bottom): Radial profile of the flux ratio between Ne~\emissiontype{IX}~He$\alpha$ 
(0.86--0.96~keV) and Si~\emissiontype{XIII}~He$\alpha$ (1.76--1.92~keV) emission. All segments are summed here.}
\label{fig:spec}
\end{center}
\end{figure}

Figure 8 clearly shows that the Ne~\emissiontype{IX}~He$\alpha$ emission extends to a radius of $\sim$8~arcmin (the outer radio shell), which is outside the main X-ray region of Kes 79 ($\sim$5~arcmin). The X-ray spectrum is extracted from a donut-shape region defined by the two circles of radii $\sim$5 and 8~arcmin (white dashed lines in the left panel of Figure~8). The spectrum is fitted by the same model as that given in subsection 3.3 and 3.4. Since statistics are limited, we fixed the NEI model to that of the whole region (Table 3). The fit is acceptable with a $\chi^2$ / d.o.f value of 324.69/304.

The best-fit spectrum and parameters are shown in Figure~9 and Table~5.  
Since the normalization ratio between the CIE and NEI components (CIE/NEI) is $\sim$100, the CIE fraction gradually increases with distance from the CCO. Thus the outer radio ring shows relative concentration of APEC, the CIE plasma. We see no 6.4 keV line from the outer radio shell region.

\begin{figure}[htbp]
\begin{center}
\includegraphics[width=7cm]{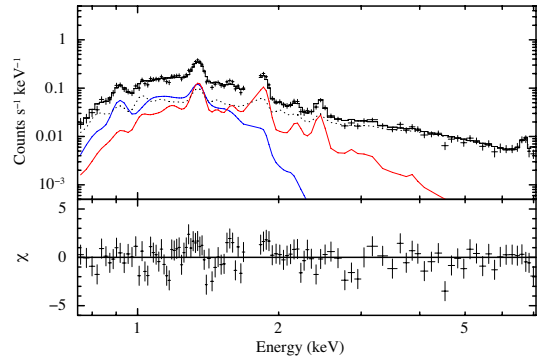}
\caption{Spectrum of the outer ring of Kes 79 (black crosses) with the best-fit models (solid lines) and the residuals (lower panel). For clarity, only the merged FI spectrum is displayed. The dotted line shows the background model. The dashed line is the NEI plasma. The red line indicates the best-fit additional CIE plasma model.}
\label{fig:spec}
\end{center}
\end{figure}
\begin{table}[htbp]
\caption{Fitting parameters of low temperature plasma}
\begin{center}
\begin{tabular}{ccc}
\hline
Component&Parameter & \\ \hline
Absorption &$N_{\mathrm{H}}$ (10$^{22}$~cm$^{-2}$) & $1.66^{+0.08}_{-0.06}$ \\
CIE&$kT$ (keV) & $0.17\pm0.02$ \\
&Abundance (solar) & 1 (fix) \\
&$EM$$^*$ ($10^{12}$~cm$^{-5}$) & 30$^{+40}_{-10}$ \\
NEI&$EM$$^*$ ($10^{12}$~cm$^{-5}$) & 0.31$\pm$0.02 \\ \hline
&$\chi^2$ / d.o.f & 324.69/304 \\ \hline
\end{tabular}
\end{center}
$^*$$\int n_{\mathrm{e}} n_{\mathrm{p}} \mathrm{d}V/4 \pi D^2$, where $n_{\mathrm{e}}$, $n_{\mathrm{p}}$, $D$  and $V$ are the electron and proton number densities (cm$^{-3}$), the distance to the SNR (cm) and emission volume (cm$^3$), respectively.\\
\label{tbl:par background}
\end{table}

\section{Discussion}
\label{sec:discussion}
\subsection {Abundances and typing of the SNR}

We have shown that the X-ray spectrum of Kes 79 is best described by two components: a CIE plasma and a multi-temperature NEI plasma. Since the CIE plasma has solar abundance, while the NEI counterpart is non-solar, the former is most naturally interpreted as interstellar medium (ISM) shocked by the SNR blast wave. 
The larger extension of the CIE plasma toward the outer radio shell also supports that the CIE plasma is blast wave shocked ISM plasma. 
The latter (NEI) would be then an ejecta  heated up by the reverse shock. 
Figure~10 shows the comparison of the overall ejecta abundance pattern of Kes 79 with available explosive nucleosynthesis models of Type Ia (CDD1: Iwamoto et al. 1999) and CC supernovae (CC-SN) (Woosley \& Weaver 1995). The measured abundance pattern broadly agrees with the $\sim$30--40~\Mo CC-SN models.
Assuming that the ejecta of Kes 79 has a roughly spherical shape with a 5~arcmin radius and distance ($D$) of 7~kpc, the X-ray emitting volume ($V$) is $1.3\times 10^{59}~d_{7}^3 f^{0.5}$~cm$^3$, where $f$ is the filling factor. 
From the normalization of the NEI emission component ($n_{\mathrm e} n_{\mathrm H} V /4\pi D^2=3.8\pm0.6\times10^{12}$~cm$^{-5}$), the total ejecta mass is estimated to be  $\sim 60~f^{0.5}~d_{7}^{5/2}~\Mo$.  This mass estimation is based on a H-dominated plasma.  Using He, C and O abundances from ejecta models of $\sim$30--40~\Mo progenitor stars, the total ejecta mass is re-estimated to be $\sim 30~f^{0.5}~d_{7}^{5/2}~\Mo$. These are reasonably consistent with a $\sim$30--40~\Mo progenitor star, taking account of the distance ambiguity.

The NEI (ejecta) plasma has a temperature structure such that lighter elements are colder than the heavier elements. This trend is found in young Ia SNRs with onion-like structure, where the heavier elements are concentrated in the inner regions, while lighter elements are in the outer layers. As is shown in Section 3.4, Kes 79 has no onion-like structure; heavier elements are rather depleted in the central core.  Therefore the observed temperature structure of this CC-SNR may be indicative of some mechanism other than that of onion-like Type Ia structure origin. 
One possibility is that the heavier elements attained higher ion temperature just after shock heating.  If the time scale of energy transfer from ion to electron is comparable or longer than the ejecta plasma age, the electrons around the heavier elements can be at higher temperature than those around the lighter elements. To justify this idea, we need detail simulations of shock heating and preceding energy transfer process from ions to electrons, which is beyond the scope of this paper.

\begin{figure}[htbp]
\begin{center}
\includegraphics[width=10cm]{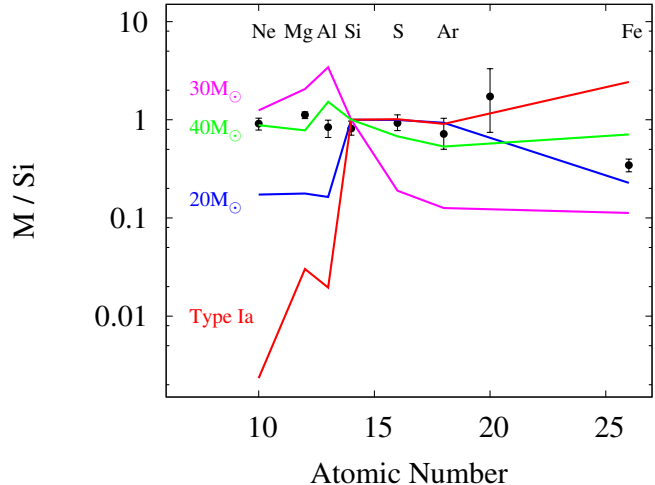}
\caption{The best-fit metal abundance pattern of the Kes 79 ejecta relative to Si as a function of atomic number. The red line shows the pattern of a Type Ia SN model (CDD1: Iwamoto et al. 1999). The blue, magenta and green lines are Type II CC-SN models with progenitor masses of 20~\Mo,  30~\Mo and 40~\Mo (S40C), respectively (Woosley \& Weaver 1995).}
\label{fig:mass}
\end{center}
\end{figure}

\subsection {X-ray emission from the outer radio ring}

Our work reveals that X-rays are extended to the outer ring of $5'$--$8'$ to beyond what was previously known. The X-ray radial profile indicates the outer X-rays are associated with the outer radio shell.  In the outer ring, the surface brightness  of the CIE (ISM) relative to NEI (ejecta) is larger than that of the inner ring and center core. 
Assuming the emitting volume of CIE to be a sphere, the inner and outer regions have a similar density $n_{\rm H} (\sim 1.0$~cm$^{-3})$ according to projection effect. The temperature of the CIE is also the same as that in the main X-ray emitting regions. 
Thus the ISM plasma is uniformly distributed in the whole SNR in terms of density and temperature.

For the angular size of $R=8$~arcmin, the radius of the outer ring is 16~pc, and the volume of the plasma is $5.4\times10^{59}$~cm$^3$. The best-fit ISM temperature of $kT_{\rm e}\sim0.2$~keV gives the expansion speed of $V_s = 3\times10^7$~cm~s$^{-1}$ from the strong shock relation assuming electron-ion temperature equilibration, $kT_{\rm e} = 3/16 \mu m_{\rm H} V_s^2$, where $\mu$ and $m_{\rm H}$ are the mean atomic mass and the mass of a hydrogen atom, respectively. Taking into account the velocity decrease in time ($t$) of $t^{-0.6}$, the dynamical age is estimated to be $\sim 2.7\times 10^4$~years from the Sedov self-similar solution (Sedov 1959). Thus Kes79 is an old-intermediate aged SNR. 

\subsection{On the Origin of 6.4~keV Line}

We discovered  a K-shell line of Fe at 6.4~keV. Since Model A gives a good fit, the inferred origin of the 6.4~keV line would be a low ionization Fe in the ejecta. 
For the progenitor type inferred from the abundance pattern, the expected Fe mass is around $M_{\mathrm {Fe}} \sim4\times10^{-2}~d_7^{2.5}$~\Mo. This mass is not inconsistent with current CC-SN models (Woosley \& Weaver 1995).  However the detection of Fe lines at $\sim6.4$~keV is very unusual for CC-SNRs (see, e.g., Yamaguchi et al. 2014).  The spatial distribution of the 6.4~keV line, no emission near the core region and enhanced emission toward the boundary of the SNR, is against the idea of the ejecta origin.

A dense molecular cloud near Kes 79 has been found using $^{13}$CO data recently (see Figure 5 of Giacani et al 2009). The $N_{\rm H}$ difference between Kes 79 and the nearby background (BG2) is $\sim 10^{22}$~cm$^{-2}$ (Table 2).
Thus, we can assume that the molecular could with $N_{\rm H}=10^{22}$~cm$^{-2}$ is just behind Kes 79.
The empirical relation between the mass and size (Scovile et al.1987) gives a mean density of H of $180 (D/40$~pc)$^{-0.9}$, which is $N_{\rm H}\sim 10^{22}$~cm$^{-2}$ for this cloud. These values are roughly consistent with our estimations from the BG2 and Kes79 spectra.

Thus, another possible origin of the 6.4~keV line is due to neutral Fe, an inner-shell ionization of the molecular cloud by the locally accelerated cosmic rays (CRs). 
Since the equivalent width (EW) of the line is larger than 2~keV, CR electrons can be excluded (e.g. Valinia et al. 2000), CR protons are the more probable candidate responsible for the 6.4~keV line production (Dogiel et al. 2011). 
The flux of the 6.4 keV line is given by 
$\frac{1}{4\pi}\sigma_{\rm 6.4~keV}\,\upsilon\,n_{\rm p}\,N_{\rm H}$, where $\sigma_{\rm 6.4~keV}$, $\upsilon$, $n_{\rm p}$, and $N_{\rm H}$ are the cross section to produce the 6.4 keV line by particles, the velocity and density of the protons, and the line-of-sight hydrogen column density, respectively. 
The 6.4 keV flux is $3.0\times10^{-6}$~cm$^{-2}$~s$^{-1}$. The cross section $\sigma_{\rm 6.4 keV}$ has a peak of 
$\sim1.3\times10^{-26}$~cm$^{2}$~hydrogen-atom$^{-1}$ at 10~MeV in solar abundance (Romo-Kr{\"o}ger 1998). We then calculate the proton energy density to be $\sim 10^2$~eV~cc$^{-1}$ by simply using a mono-energetic distribution. This is about 100 times larger than the canonical value $\sim1$~eV~cc$^{-1}$ in the ISM, and hence implies a production site near the SNR whose forward shock is accelerating the protons responsible for the ionization.  A similar connection between the 6.4 keV line and $\sim10$~MeV protons has observationally confirmed by Nobukawa et al. (2015).

The CR protons also produce the majority of the observed gamma-rays through the $\pi^0$-decay mechanism (e.g., Auchettl \etal\ (2014), Lee \etal\ (2015)). Although we cannot unveil the spatial distribution of this line emission beyond the angular resolution of \textit{Suzaku}, the detection of the 6.4~keV line has strong implications on the local environment, and possibly for the GeV gamma-ray emission.   
We note that the effect of CR energy losses is non-negligible in the MeV energy range in a dense molecular environment. On the other hand, the $\pi^0$-decay gamma-rays are produced by protons of energies above the pion production threshold and thus are not much affected by Coulomb losses.  Therefore, the larger energy content inferred from the GeV emission (Auchettl \etal\ 2014) than from the 6.4~keV line (this work) is not surprising. 

Since the diffusion length of $\sim10$~MeV protons is $\sim1$~pc, the shock front of the blast wave (site of the $\sim10$~MeV proton production) should be in contact with the molecular clouds (site of the 6.4~keV line production). This situation is different from the larger diffusion  length of GeV proton and resulting GeV gamma-ray emission, and hence may also relax the apparent energy deficiency of the MeV protons compared to the GeV protons; the GeV gamma-ray region would have an extension larger than that of the 6.4~keV line region.

\section{Conclusions}

We have carefully studied the Galactic SNR Kes 79 using \textit{Suzaku} data and a reliable background model of GRXE near the SNR.
A 6.4~keV K$\alpha$ line from nearly neutral Fe is discovered, and is found to associate spatially with dense molecular clouds interacting with Kes 79.
A number of other interesting new results have also been obtained and are summarized below:

\begin{itemize}
\item The X-ray spectrum of Kes 79 is well described by a multi-temperature NEI plasma plus a low-temperature CIE plasma, in addition to a $\sim$6.4~keV K$\alpha$ line from Fe~\emissiontype{I}. 

\item The abundance pattern and mass of the NEI plasma indicate that the SNR originates from the core-collapse of a $\sim$30--40~\Mo progenitor.

\item The CIE plasma has solar abundances and is spatially extended beyond the super-solar NEI plasma, to the outer radio ring. The CIE plasma is hence most probably ambient gas run over and heated by the forward shock.

\item The $\sim$6.4~keV line feature can be produced either from the unshocked ejecta or a MC interacting with low-energy CR protons accelerated locally by the SNR shock. However, its spatial association with a MC points to the latter possibility. 

\end{itemize}

\bigskip
The authors express sincere thanks to all the \textit{Suzaku} team members. We also thank Poshak Gandhi for carefully reading our manuscript.
TS is supported by the Japan Society for the Promotion of Science (JSPS) Research Fellowship for Young Scientists.
SHL acknowledges support by the JAXA International Top Young Fellowship.


\end{document}